\documentstyle[preprint,aps,epsf]{revtex}
\begin{document}
\draft

\preprint{
\noindent
\begin{minipage}[t]{3in}
\begin{flushright}
SLAC--PUB--95--7063 \\
December 1995
\end{flushright}
\end{minipage}
}

\title{Renormalization Scale Setting for Evolution Equation of
Non-Singlet Structure Functions and Their Moments}

\author{Wing Kai Wong
\thanks{Work supported by the Department of Energy, contract
DE--AC03--76SF00515.}
}

\address{Stanford Linear Accelerator Center, \\
Stanford University, Stanford, California 94309}

\maketitle

\begin{abstract}

We use the BLM procedure to eliminate the renormalization scale
ambiguity in the evolution equation for
the non-singlet deep-inelastic structure
function $F_2^{\text NS}(x,Q).$
The scale of the QCD coupling in the
$\overline{\text{MS}}$ scheme has the form
$Q^*(x) = Q (1-x)^{1/2} / x~f(x) $,
where $x$ is the Bjorken variable and $f(x)$ is a smoothly varying
function bounded between 0.30 to 0.45.  
Equivalently, the evolution of the $n$th moment of the structure
function should contain an effective $\Lambda_{\text{QCD}}$ pattern, with
$\Lambda_n \sim n^{1/2}$.  This variation of $\Lambda_n$ agrees with
experimental data.

\centerline{(Submitted to Physical Review {\bf D})}

\end{abstract}
 


\newpage 

 
\section{INTRODUCTION}

One of the most serious problems preventing precise empirical tests of
QCD is the ambiguity of the renormalization scale $\mu$ of
perturbative predictions.  Formally, any physical quantity should be
$\mu$ independent.  However, in practice, spurious $\mu$ dependence
appears as one can only have finite-order predictions in a
perturbative theory.  Unless one specifies the argument of the
coupling $\alpha_s(\mu)$ in the truncated predictions, the range of
theoretical uncertainty can be much larger than the
experimental error.  Although the uncertainty of the prediction
contributed from the renormalization scale ambiguity is expected to
reduce as the order of the prediction gets higher, we still need a
rational way of choosing a value for $\mu$ such that the truncated
prediction will approximate the true prediction as close as possible.

A conventional way of setting the renormalization scale in
perturbation theory is setting $\mu$ to be the momentum transfer or
energy scale $Q$ of the system (provided there is only one energy
scale); this eliminates large logarithmic terms $\ln(Q/\mu)$ and so
gives a more convergent series.  However, often an order-1 variation
of $\mu/Q$ leads to a significant uncertainty in the prediction of
perturbative QCD.

In multi-scale processes, the problem of scale setting becomes
compounded, since the choice for $\mu$ can depend on any combination
of the available physical scales.  An example of this is the proper
scale for the running coupling constant that appears in the DGLAP
evolution equation for the deep-inelastic structure function.  In
addition to the momentum transfer $Q$ of the lepton, the physical
scale can also depend on the Bjorken ratio $x =Q^2/(2p \cdot q).$
Equivalently, the scale controlling the evolution of each moment
$M_n(Q)$ of the structure function can depend on both $Q$ and $n$.

Collins\cite {Collins}, Neubert \cite{Neubert}, and Lepage and
Mackenzie \cite{lm} have emphasized that the renormalization scale
should not be fixed by an {\it ad hoc} procedure but rather, should be
determined systematically as the mean virtuality of the underlying
physical subprocess.  From this point of view, the choice of the
renormalization scale $\mu$ for a particular prediction to a certain
order thus depends on the specific experimental measure and the
truncation order.

In this paper, we discuss how to obtain the optimal scale for the
structure function evolution using the BLM (Brodsky-Lepage-Mackenzie)
\cite{BLM} scale-fixing procedure.  In this procedure, the vacuum
polarization diagrams that contribute to the non-zero QCD beta
function are resummed into the running coupling.  More technically, we
absorb into the running coupling all factors of the number of flavors
$n_f$ that appear in the coefficients of the perturbative expansion.
This criterion automatically sets the scale to a value that reflects
the average gluon virtuality of the subprocess.

On the one hand, we can calculate the evolution of the moment by
fixing the scale as a function of $Q$ and $x$ before integrating over
$x$ to obtain the moments; on the other hand, we can set the scale
from the moment-evolution equation for each $n$. We give the
appropriate scale for the moment-evolution equation for each $n$ and
show that the two procedures give the same result to the order that we
consider.

\section{ THE BLM SCALE-FIXING PROCEDURE}
 
The crucial idea of BLM procedure \cite{BLM,csr} is that, in the
$\alpha_V$ scheme [$\alpha_V$ is defined through the heavy quark
potential $V(Q^2)$ by $V(Q^2) = -4\pi C_F\alpha_V(Q)/Q^2$], the scale
must be chosen to absorb all the vacuum polarization (non-zero beta
function) contributions into the running coupling constant.  Thus one
has to choose the argument of the coupling constant in each order of
perturbation theory so that there is no $n_f$ dependence in the
coefficients of the coupling constants (those light-by-light
contributions that are not associated with renormalization are not
resummed).  This is a good choice of scale since all the
vacuum-polarization contributions are then automatically summed to all
orders for any finite-order prediction.  For a next-to-leading order
(NLO) perturbative prediction in {\em any} renormalization scheme, the
procedure simply translates to choosing the scale such that there is
no $n_f$ dependence in the NLO coefficient of the coupling constant,
by the transitivity of the procedure to this order \cite{csr}.

To show explicitly how to set the scale for an NLO perturbative QCD
prediction, consider a prediction in any scheme of the following form,
\begin{equation}
\rho = \rho_0 \alpha(\mu) \left\{ 1 +
\left[A(\mu)n_f + B(\mu)\right] {\alpha(\mu) \over \pi}
\right\}
+O(\alpha^3),
\label{e1}
\end{equation}
where $\mu$ is the renormalization scale.  The parameters $\rho$,
$\rho_0$, $A$ and $B$ are all in general dependent on one or more
physical scales such as center-of-mass energy, momentum transfer and
Bjorken parameter $x$.  The $\mu$ dependence of the coefficients $A(\mu)$
and $B(\mu)$, fixed by the renormalization group
\begin{equation}
\frac{d\rho}{d\mu} = 0 + O(\alpha^3)
\label{e2}
\end{equation}
are of the form
\begin{eqnarray}
\nonumber
A(\mu) &=& A' - \frac{1}{3}\ln(\mu) \label{Amu}\\
B(\mu) &=& B' + \frac{11}{6}C_A\ln(\mu) \label{Bmu}\ , 
\end{eqnarray}
where $A'$ and $B'$ are $\mu$ independent.

We apply the BLM procedure to eliminate the $n_f$ and $\beta_0$ dependence in
Eq. (\ref{e1}). We choose the renormalization scale $\mu$ to be

\begin{equation}
\label{blmscale}
Q^* = \mu\exp(3A(\mu)),
\end{equation}
so that 
\begin{equation}
A(Q^*) = 0
\end{equation}
[note that the right-hand side of Eq. (\ref{blmscale}) is $\mu$ independent by
Eq. (\ref{Amu})].

Using the BLM scale, the perturbative QCD prediction becomes
\begin{equation}
\rho = \rho_0 \alpha(Q^*)
\left[ 1 + B(Q^*) {\alpha (Q^*) \over \pi}
\right]
+O(\alpha^3).
\end{equation}
Rewrite it in terms of the original coefficients $A(\mu)$ and $B(\mu)$ using
Eq. (\ref{Bmu}),
\begin{equation}
\rho = \rho_0 \alpha(Q^*)
\left\{1 + \left[ \frac{11}{2}C_AA(\mu)+B(\mu)\right] {\alpha (Q^*) \over \pi}
\right\}
+O(\alpha^3).
\label{scalefixed}
\end{equation}

This is the scale-fixed perturbative QCD prediction.
With the chosen scale $Q^*$, all vacuum polarization is resummed into
the running coupling constant, with the coefficient of $\alpha^2$
independent of $n_f$.

In short, by comparing Eq. (\ref{e1}) and Eq. (\ref{scalefixed}), 
setting the scale for an observable is simply equivalent to replacing $n_f$ by
$\frac{11}{2}C_A$, and using $Q^*$ as the argument of the coupling
constant ($\mu$ dependences in $A(\mu)$ and $B(\mu)$ cancel).

For a prediction with $\mu$ preset to any energy scale $Q$, as is
usually given in the literature, the procedure for resetting
the scale appropriately is identical to the one presented above except with
$Q$ replacing $\mu$.

As we have seen, to obtain the renormalization scale only the $n_f$
term of the NLO result is needed.  Therefore, one can improve a
leading-order result by setting an appropriate scale without
calculating the full NLO prediction.  Extensions of the method to
higher orders are given in \cite{csr}, \cite{gk}, and \cite{bb}.

\section{ SCALE SETTING OF THE STRUCTURE-FUNCTION EVOLUTION EQUATION}
 
The non-singlet, structure-function evolution equation is \cite{ap}
 
\begin{equation}
Q^2 \frac{\partial {\cal F}^{\text{NS}}_2}{\partial Q^2}(x,Q) =
\int_{x}^{1}\frac{dy}{y}
\left[P_{qq}\left(\frac{x}{y},\alpha(Q)\right)
\pm P_{\bar q q}\left(\frac{x}{y},\alpha(Q)\right)\right]
{\cal F}^{\text{NS}}_2(y,Q),
\label{structure function evolution equation}
\end{equation}
where, in terms of experimental quantities,
\begin{equation}
{\cal F}^{\text{NS}}_2 = \left\{
\begin{array}{llll}
(F^{\text{eP}}_2-F^{\text{eN}}_2)/x
&\text{for} +\ ,  \\
(F^{\text{$\bar \nu$P}}_2-F^{\text{$\nu$P}}_2)/x
&\text{for} - \ ,
\end{array}
\right.
\end{equation}
and $P_{qq} \pm P_{\bar q q}$ are the non-singlet evolution kernels, for
which the NLO 
perturbative predictions in $\overline{\text{MS}}$ scheme with
the renormalization scale preset to momentum transfer $Q$ are
\cite{frs}, \cite{cfp}, \cite{flk}:

\begin{mathletters}
\label{kernels}
\begin{eqnarray}
P_{qq}(x, \alpha_{\overline{\text{MS}}}(Q))
          & = & \frac{\alpha_{\overline{\text{MS}}}(Q)}{2\pi}
                 C_F \left(\frac{1+x^2}{1-x}\right)_+
		+\left(
			\frac{\alpha_{\overline{\text{MS}}}(Q)}{2\pi}
		\right)^2
		\left\{
                	C_F^2 \left[
					-2 \left(\frac{1+x^2}{1-x}\right)
                			   	\ln x \ln(1-x)
			      \right.
                \right. \nonumber \\
          && -   	      \left.    5(1-x)
					- \left(\frac{3}{1-x}+2x\right)\ln x
					- \frac{1}{2}(1+x) \ln^2 x
			      \right] \nonumber \\
          && +  \frac{1}{2}C_FC_A\left[
					\left(\frac{1+x^2}{1-x}\right)
                  			\left(\ln^2 x
					  - \frac{11}{3} \ln \frac{1-x}{x^2}
					  + \frac{367}{16}
					  - \frac{\pi^2}{3}
					\right)
				\right. \nonumber \\
          && +                  \left.	2(1+x)\ln x
					+ \frac{61}{12}
					- \frac{215}{12} x
				\right] \nonumber \\
          && +  \left. \frac{2}{3}C_FT\left[
					\left(\frac{1+x^2}{1-x}\right)
                			\left(\ln\frac{1-x}{x^2}
						-\frac{29}{12}
					\right)
					+ \frac{1}{4}
					+\frac{13}{4}x
				\right]
		\right\}_+ \nonumber \\
          && +  \delta (1-x)
                \int_0^1dxP_{\bar qq}(x,\alpha_{\overline{\text{MS}}}(Q))
+O(\alpha_s^3).
\label{Pqq}
\end{eqnarray}
and
\begin{eqnarray}
P_{\bar q q}(x, \alpha_{\overline{\text{MS}}}(Q))
&=& \left(\frac{\alpha_{\overline{\text{MS}}}(Q)}{2\pi}\right)^2
(C_F-\frac{1}{2}C_A)C_F
\left[2(1+x)\ln x +4(1-x) \right. \nonumber \\
&&+ \left. \frac{1+x^2}{1+x}\left(\ln ^2x-4\ln x \ln (1+x) - 4
\text{Li}_2(-x)
-\frac{\pi^2}{3}\right)\right], 	
\label{Pbarqq}
\end{eqnarray}
\end{mathletters}
with
\begin{eqnarray}
T     &=& n_f/2 \nonumber \\
C_A   &=& 3     \nonumber \\
C_F   &=& 4/3   \nonumber \\
\text{Li}_2(x) &=& - \int_0^x\frac{dy}{y}\ln (1-y).
\end{eqnarray}
The $+$~distribution
notation serves as a regulator defined as
\begin{equation}
\int_z^1dx f(x) \left[\frac{g(x)}{1-x}\right]_+
= \int_z^1dx\frac{(f(x)-f(1))g(x)}{1-x} - f(1)\int_0^zdx\frac{g(x)}{1-x}.
\end{equation}
 
Since all the coupling constants reside in the evolution kernels,
we set the scale
for the structure-function evolution equation by setting the
scale for the kernels.
When the kernels $P_{qq} \pm P_{\bar q q}$
have been recast in the form of Eq. (\ref{e1}),
the coefficient of $n_f$ can be obtained as
\begin{equation}
A(x) = \frac{1}{6} \left[
                        \ln \left( \frac{1-x}{x^2} \right)
                      - \frac{29}{12}
                      + \left( \frac{1-x}{1+x^2} \right)
                        \left( \frac{1}{4} + \frac{13}{4} x \right)
               \right].
\label{Ax}
\end{equation}
 
\begin{figure}
\begin{center}
\leavevmode
{\epsfbox{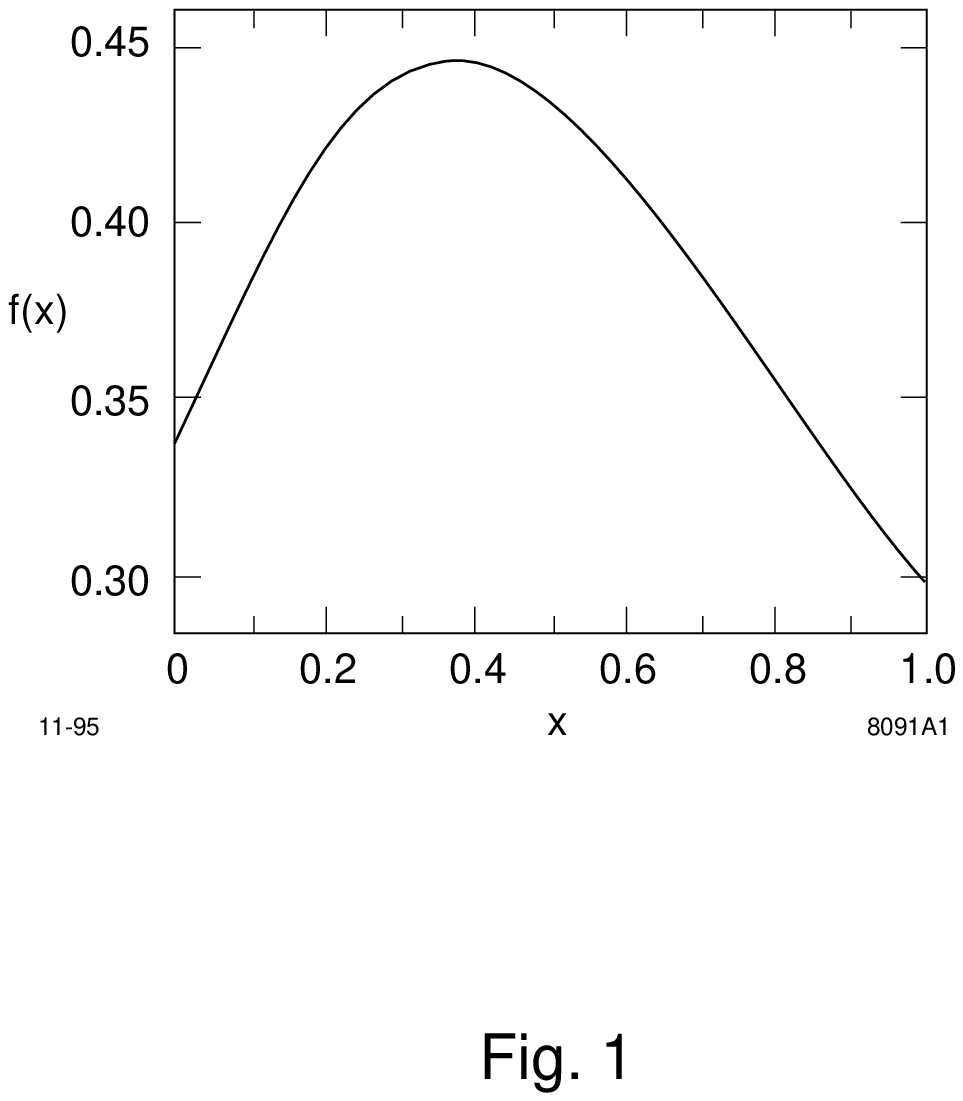}}
\end{center}
\caption{Plot of $f(x)$ against $x$.}
\label{f1}
\end{figure}

Using Eq. (\ref{blmscale}), BLM scale-fixing procedure gives for the
BLM scale
\begin{equation}
Q^*(x) = Q \exp(3A(x)) = Q \frac{(1-x)^{1/2}}{x}~f(x),
\label{kernelscale}
\end{equation}
where
\begin{equation}
f(x) = \exp \left[
\frac{1}{2}
\left(\frac{1-x}{1+x^2}\right)
\left(\frac{1}{4} +\frac{13}{4}x\right)
-\frac{29}{24}
\right] 
\end{equation}
is a smoothly varying function bounded between 0.30 and 0.45 as
shown in Fig. \ref{f1}.
 
Note that the above manipulation is done inside the $+$ distribution
notation.  It is because after setting the scale to be dependent on
$x$, the coupling constants must also be included in the $+$
distribution notation in order to preserve the Adler sum rule.
 
Finally, the scale-fixed evolution kernels can be obtained
by rewriting Eq. (\ref{kernels}) in the form of Eq. (\ref{scalefixed}), using
the BLM scale given by Eq. (\ref{kernelscale}):
\begin{mathletters}
\label{scalefixedP}
\begin{eqnarray}
\tilde{P}_{qq}(x, \alpha_{\overline{\text{MS}}}(Q^*(x)))
          & = &  \left\{ \frac{\alpha_{\overline{\text{MS}}}(Q^*(x))}{2\pi}
                 C_F \left(\frac{1+x^2}{1-x}\right)
		+\left(
			\frac{\alpha_{\overline{\text{MS}}}(Q^*(x))}{2\pi}
		\right)^2
		\right. \nonumber \\
	  && \times 
		\left\{
                	C_F^2 \left[
					-2 \left(\frac{1+x^2}{1-x}\right)
                			   	\ln x \ln(1-x)
			      \right.
                \right. \nonumber \\
          && -   	      \left.    5(1-x)
					- \left(\frac{3}{1-x}+2x\right)\ln x
					- \frac{1}{2}(1+x) \ln^2 x
			      \right] \nonumber \\
          && +  \left. \left. \frac{1}{2}C_FC_A\left[
					\left(\frac{1+x^2}{1-x}\right)
                  			\left(\ln^2 x
					  + \frac{2027}{144}
					  - \frac{\pi^2}{3}
					\right) 
		\right. \right. \right. \nonumber \\
	  && +  \left. \left. \left. 2(1+x)\ln x - 6x + 6
				\right] \right\} \right\}_+ \nonumber \\
          && +  \ \delta (1-x)
                \int_0^1dx\tilde{P}_{\bar q q}(x,
		\alpha_{\overline{\text{MS}}}(Q^*(x)))
               +O(\alpha_s^3),
\label{scalefixedPqq}
\end{eqnarray}
and
\begin{eqnarray}
\tilde{P}_{\bar q q}(x, \alpha_{\overline{\text{MS}}}(Q^*(x)))
&=& \left(\frac{\alpha_{\overline{\text{MS}}}(Q^*(x))}{2\pi}\right)^2
		(C_F-\frac{1}{2}C_A)C_F
\left[2(1+x)\ln x +4(1-x) \right. \nonumber \\
&&+ \left. \frac{1+x^2}{1+x}\left(\ln ^2x-4\ln x \ln (1+x) - 4
\text{Li}_2(-x)
-\frac{\pi^2}{3}\right)\right]\ .	
\label{scalefixedPbarqq}
\end{eqnarray}
\end{mathletters}
 
The scale fixed-structure function evolution equation is then given by
\begin{equation}
Q^2 \frac{\partial {\cal F}^{NS}_2}{\partial Q^2}(x,Q) =
\int_{x}^{1}\frac{dy}{y}
\left[\tilde{P}_{qq}\left(\frac{x}{y},\alpha_{\overline{\text{MS}}}
(Q^*(\frac{x}{y}))\right) \pm
\tilde{P}_{\bar q q}\left(\frac{x}{y},\alpha_{\overline{\text{MS}}}
(Q^*(\frac{x}{y}))\right)\right]
{\cal F}^{\text{NS}}_2(y,Q).
\label{scale fixed structure function evolution equation}
\end{equation}

This equation can also be written in terms of $\alpha_V$ using the
relation \cite{BLM} 
\begin{equation}
\alpha_{\overline{\text{MS}}}(\mu) =
\alpha_{V}(\mu\exp(5/6))(1+2\alpha_V/\pi+\ldots).
\label{ms2v}
\end{equation} 

\section{ VERIFICATION OF THE SCALE}
 
A good way to see the validity of the scale set for the
structure-function evolution equation is to apply it to moment
evolution and compare the results with experimental data. The moment
analysis of the Fermilab muon deep-inelastic scattering data
\cite{data} showed an interesting variation of the effective
$\Lambda_{\text{QCD}}$ with $n$, both at leading and next-to-leading
order.  Here we would like to demonstrate that this variation can be
readily explained by the choice of the renormalization scale even at
leading order.

The $n$th ordinary moment of the non-singlet structure function is defined as:
\begin{equation}
M_n(Q^2) = \int_0^1dxx^{n-1}{\cal F}^{\text{NS}}_2(x,Q^2).
\end{equation}
The Nachtman moments will give the same result up to higher twist terms.

Taking the $n$th moment of the structure function evolution equation
(Eq. (\ref{scale fixed structure function evolution equation})),
interchanging the order of integration on the right-hand side and rearranging,
one can obtain the equation of evolution for the moment:
\begin{equation}
\frac{\partial\ln M_n(Q^2)}{\partial\ln Q^2} = \int_0^1dxx^{n-1}
\left(\tilde{P}_{qq}(x, \alpha(Q^*(x)))
\pm \tilde{P}_{\bar q q}(x, \alpha(Q^*(x))\right),
\label{moment evolution}
\end{equation}
which is just the $n$th moment of the evolution kernels.
 
Let us consider only the leading order prediction of the kernel
function in Eq. (\ref{scalefixedPqq}), then Eq. (\ref{moment evolution})
becomes
\begin{equation}
\frac{\partial\ln M_n(Q^2)}{\partial\ln Q^2} = \int_0^1\left(
\frac{\alpha(Q^*(x))}{2\pi}
           P_{qq}^0(x)                                  \right)_+
x^{n-1}dx,
\label{e13}
\end{equation}
here $Q^*(x)$ is given by Eq. (\ref{kernelscale}),
and
\begin{equation}
P_{qq}^0(x) = C_F\frac{1+x^2}{1-x}
\end{equation}
the leading order coefficient.
 
The crucial difference between the conventional calculation, that is, using
the momentum transfer $Q$ as the renormalization scale, and the
calculation here, namely using the BLM scale, is that the
argument of the coupling constant in Eq. (\ref{e13}) is now $x$ dependent.
 
To see the effect of the scale setting on the leading-order analysis,
expand $\alpha(Q^*)$ in powers of $\beta_0\alpha(Q)$
(similar results can be obtained without the expansion), that is,
\begin{equation}
\alpha(Q^*(x)) = \alpha(Q)\left\{1+
\beta_0\ln \left(\frac{Q}{Q^*(x)}\right) \frac{\alpha(Q)}{2\pi} + \ldots
\right\},
\label{alpha_expansion}
\end{equation}
 
then Eq. (\ref{e13}) becomes
\begin{equation}
\frac{\partial\ln M_n(Q^2)}{\partial\ln Q^2} =
\frac{\alpha(Q)}{2\pi}A_n\left[
1+\beta_0B_n\frac{\alpha(Q)}{2\pi} + \ldots
\right],
\label{e15}
\end{equation}
 
where
\begin{eqnarray}
A_n &=& \int_0^1dxx^{n-1} P_{qq}^0(x)_+, \text{and} \\
B_n &=& \frac{1}{A_n}\int_0^1dxx^{n-1} \left[ P_{qq}^0(x)
\ln \left( \frac{Q}{Q^*(x)}\right)
\right]_+.
\end{eqnarray}

Now Eq. (\ref{e15}) can be integrated with respect to $\ln Q^2$
by using, to LO (to compare with LO experimental analysis), 
\begin{equation}
\alpha(Q) = \frac{4\pi}{\beta_0 \ln (Q^2/ \Lambda^2)},
\label{coupling constant}
\end{equation}
and the solution for $M_n(Q)$ is then
\begin{equation}
M_n(Q)^{-1/d_n} =C_n\ln(Q^2/\Lambda^2)
\exp\left[-\frac{2B_n}{\ln(Q^2/\Lambda^2)}\right],
\end{equation}
where $d_n = -2A_n/\beta_0$ and $C_n$ are constants.

This is our prediction for the evolution of the moments.
 
For experimental data taken at large $Q$, the exponential term can
be expanded and we have an equation linear in $\ln(Q^2/\Lambda^2)$:
\begin{equation}
M_n(Q)^{-1/d_n} =C_n\left[\ln(Q^2/\Lambda^2)- 2B_n\right].
\label{e20}
\end{equation}
 
Had we set the scale to $Q$, we would have had
\begin{equation}
M_n(Q)^{-1/d_n} =C_n\left[\ln(Q^2/{\Lambda_n}^2)\right],
\end{equation}
 where $\Lambda_n$ is an ``effective'' $\Lambda$ at LO and
will depend on $n$ as predicted by Eq. (\ref{e20}) as
\begin{equation}
\Lambda_n = \Lambda/ \exp (-B_n).
\end{equation}

\begin{figure}
\begin{center}
\leavevmode
{\epsfbox{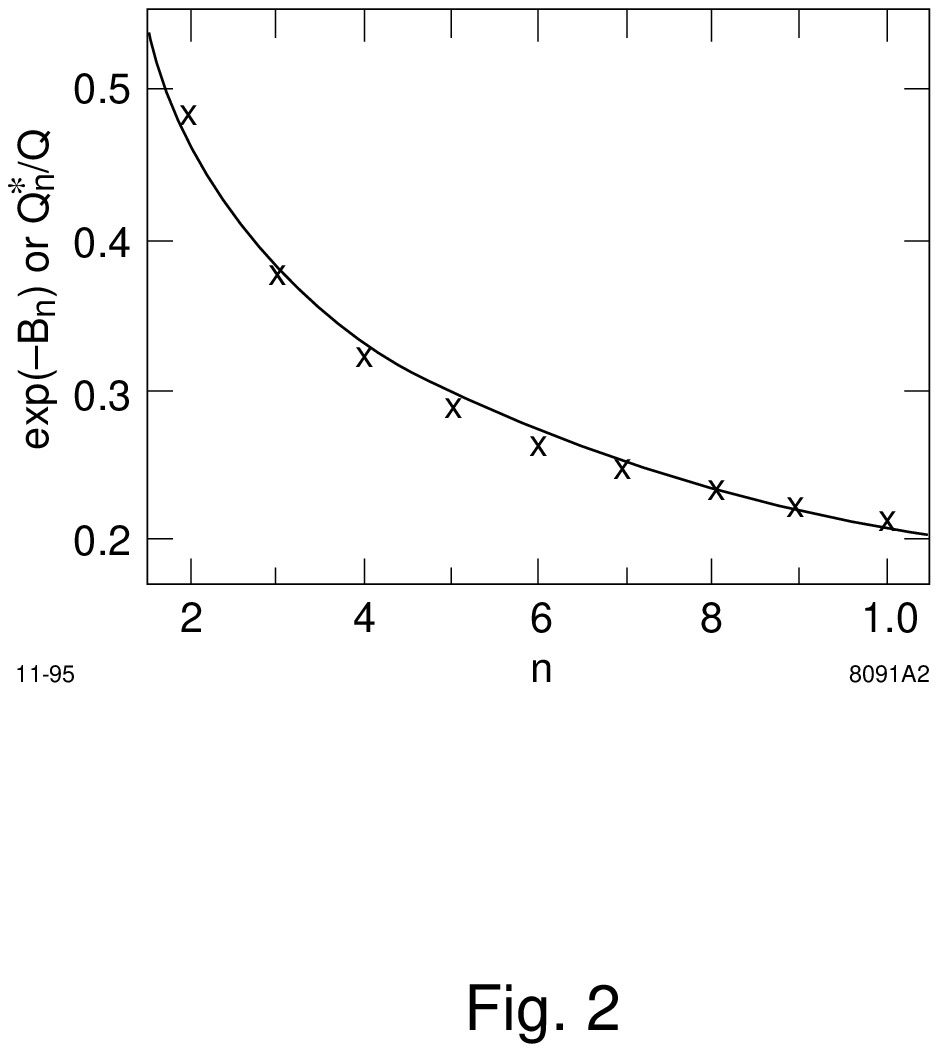}}
\end{center}
\caption{Plot of $\exp(-B_n)$ (or $Q^*_n/Q$) against $n$. The fitted
curve has the form $\exp(-B_n) = 0.656/n^{1/2}$.}
\label{f3}
\end{figure}

Values of $\exp(-B_n)$ for $n$ from 2 to 10 are plotted in
Fig. \ref{f3}; these are well-described by the form 
\begin{equation}
\exp(-B_n) = 0.656/n^{1/2}.
\end{equation}

Therefore, we expect the dependence of $\Lambda_n$ on $n$ to be
\begin{equation}
\Lambda_n = \frac{n^{1/2}}{0.656} \Lambda.
\label{e21}
\end{equation}

\begin{figure}
\begin{center}
\leavevmode
{\epsfbox{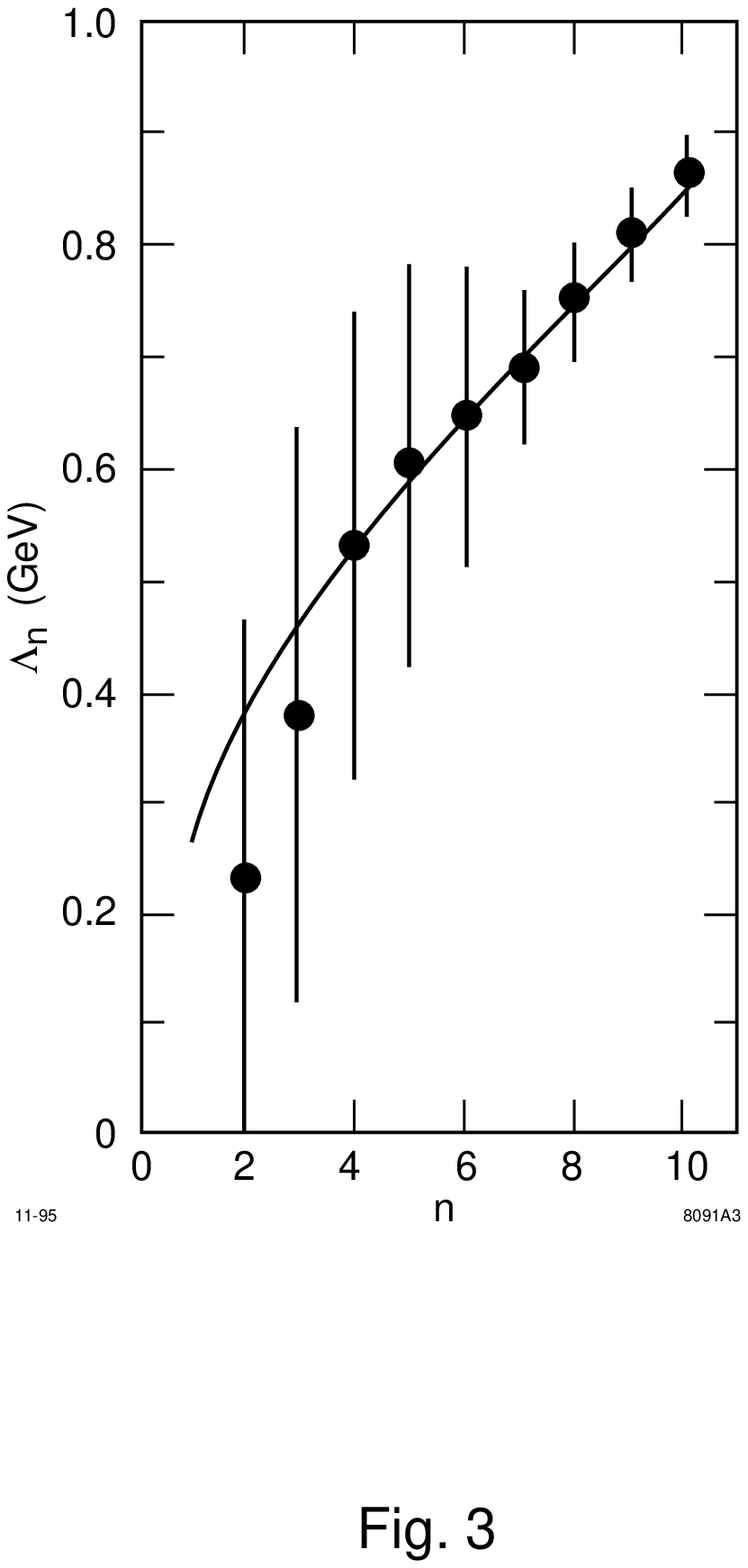}}
\end{center}
\caption{Plot of $\Lambda_n$ against $n$, after B. A. Gordon et
al.}
\label{f4}
\end{figure}

In fact, the data of Gordon {\it et al} does show the predicted
variation.  In Fig. \ref{f4} we show the results extracted for $\Lambda_n$
against $n$.  A fit to Eq. (\ref{e21})
yields $\Lambda_{\overline{\text{MS}}} = 170 \pm 120$ MeV, where the error is
estimated from higher-order terms in Eq. (\ref{alpha_expansion}).

\section{ SCALE SETTING FOR THE MOMENT-EVOLUTION EQUATION}
 
The BLM scale fixing procedure can also be applied to the
moment-evolution equation {\em after} the moment integration has been
carried out, with evolution kernels given by Eq. (\ref{kernels}).
When the resulting equation is written in the form of Eq. (\ref{e1}),
the coefficient of $n_f$ is given by
\begin{equation}
A =\frac{ \int_0^1dxx^{n-1}\left(P_{qq}^0A(x)\right)_+}{A_n},
\end{equation}
where $A(x)$ is given by Eq. (\ref{Ax}).
This is related to our previously defined $B_n$ as
\begin{equation}
A = -\frac{B_n}{3}.
\end{equation}
 
Again, using Eq. (\ref{blmscale}), the BLM scale for the moment-evolution
equation is
\begin{equation}
Q^*_n = Q \exp(-B_n).
\end{equation}
Now the scale is $x$ independent but $n$ dependent.
 
The scale for the moment-evolution equation can also be read  from
Fig. \ref{f3} and has  a fit of
$Q^*_n/Q = 0.656/n^{1/2}$.  The $n^{1/2}$ behavior for large $n$ was
also predicted in \cite{BLM}.
 
Keeping only the leading-order term of $P_{qq}$, the moment-evolution
equation can be written  analogous to Eq. (\ref{e15}) as
\begin{equation}
\frac{\partial\ln M_n(Q^2)}{\partial\ln Q^2}
= \frac{\alpha(Q^*_n)}{2\pi}A_n.
\end{equation}
 
Again, it can be solved by rewriting the coupling constant using
Eq. (\ref{coupling constant}) to be
\begin{eqnarray}
M_n(Q)^{-1/d_n} & = & C_n\ln({Q^*_n}^2/\Lambda^2) \nonumber \\
                & = & C_n\left[\ln(Q^2/\Lambda^2)- 2B_n\right],
\end{eqnarray}
which is simply Eq. (\ref{e20}).  Actually, the commutativity of
scale setting and moment integration works at any higher order,
because a choice of the renormalization scale can only affect the
result by an order higher than that of the calculation.
 
\section{CONCLUSION}

We have applied the BLM procedure to set the scale in the evolution
equations for the non-singlet structure functions and their moments in
deep-inelastic scattering.  The variation of the effective
$\Lambda$ obtained from the $n$th-order moment-evolution data
is explained by this method, and an unambiguous value of
$\Lambda_{\overline{\text{MS}}}$ is obtained.

\acknowledgements
It is a pleasure to thank Stanley Brodsky, Lance Dixon, Hung Jung Lu,
Michael Peskin, and Arvind Rajaraman for helpful discussions.


%
%
 
\end{document}